# Heat conductivity and Dufour effects in the weakly ionized, magnetized and kappa-distributed plasma

Haijuan Xia | Jiulin Du

*Department of Physics, School of Science, Tianjin University, Tianjin 300072, China*

**ABSTRACT** By using the generalized Boltzmann equation of transport and the first-order approximation of Chapman-Enskog expansion on the $\kappa$-distribution function, we study the thermal conductivity and Dufour effects in the weakly ionized and magnetized plasma. We show that in the $\kappa$-distributed plasma, the thermal conductivity and Dufour coefficient are significantly different from those in the Maxell-distributed plasma, and the transverse thermal conductivity and Dufour coefficient in the $\kappa$-distributed plasma are generally greater than those in the Maxwell-distributed plasma, and the Righi-Leduc coefficient and Hall Dufour coefficient in the $\kappa$-distributed plasma are also generally greater than those in the Maxwell-distributed plasma.

## 1 | INTRODUCTION

Non-Maxwellian distributions and/or power-law distributions have been observed and studied extensively in many complex systems, such as astrophysical self-gravitating systems, space and astrophysical plasmas, high energy physical systems, biological and chemical systems, analysis of time series and images etc. It has been found that the velocity distributions of particles in the plasmas with long-range interactions and long-range correlations are generally far from the Maxwell distribution and with long energetic tails. Especially in the astrophysical and space plasmas, early in 1968, Vasyliunas introduced the power-law kappa-distribution by simulating the velocity distribution of energetic electrons in the electron spectrum observed in the plasma sheath of the magnetosphere,[1-4] which can be expressed as

$$f_\kappa(\varepsilon) = nB_\kappa \left[ 1 + \frac{1}{\kappa}\frac{\varepsilon}{\varepsilon_0} \right]^{-(\kappa+1)}, \tag{1}$$

with the normalized $\kappa$-constant,

$$B_\kappa = \left[ \frac{2\pi k_B T}{m}\left(\kappa - \frac{3}{2}\right) \right]^{-3/2} \frac{\Gamma(\kappa+1)}{\Gamma(\kappa - \frac{1}{2})}.$$

In the $\kappa$–distribution (1), $n$ is density, $k_B$ is Boltzmann constant, $T$ is temperature, $\varepsilon = m(\mathbf{v}\text{-}\mathbf{u})^2/2$ is the kinetic energy with speed $\mathbf{v}$, particle mass $m$ and the bulk velocity $\mathbf{u}$ of the plasma, $\varepsilon_0 = m w_\kappa^2/2$ is a characteristic energy with the speed $w_\kappa$ related to the most probable speed $w_0$ of a Maxwellian velocity distribution by

$$w_\kappa = w_0 \left[ \left(\kappa - \frac{3}{2}\right)\kappa^{-1} \right]^{1/2}, \quad \kappa > \frac{3}{2}. \tag{2}$$

The $\kappa$-distribution (1) is a power-law velocity distribution, where the parameter $\kappa \ne \infty$ describes the degree of deviation away from a thermal equilibrium state in the nonequilibrium complex plasma. Only when we take the limit $\kappa \to \infty$, the $\kappa$-distribution (1) becomes a Maxwellian velocity distribution. Recently, a relation of the $\kappa$-parameter for a nonequilibrium complex plasma with the magnetic field **B** was derived. This relation can be expressed [5] as

$$(\kappa - \frac{3}{2})k_B \nabla T = e[-\nabla \varphi_c + c^{-1} \mathbf{u} \times \mathbf{B}], \quad (3)$$

where $\varphi_c$ is the Coulomb potential, $e$ is the electron charge and $c$ is the light speed. In Eq.(3), it is clear that the parameter $\kappa \neq \infty$ if and only if $\nabla T \neq 0$. Thus, the velocity $\kappa$-distribution (1) represents the statistical characteristic of the nonequilibrium complex plasmas, where the Coulomb long-range interactions and magnetic field play significant role. And therefore, to some extent, the $\kappa$-parameter can be calculated theoretically for given plasma.

Nonequilibrium and magnetized plasma is a complex multi-body system, in which the magnetic field is of great significance for the study of plasma transport properties. The plasma behavior depends not only on the interactions between plasma particles, but also on the interactions between particles and the magnetic field. In the absence of magnetic field, the particle motion between two collisions is a linear motion. When the magnetic field is present, the particles are subjected to Lorentz force. As we know, Lorentz force is always perpendicular to the particle's velocity, so magnetic field does not work on moving charged particles. Therefore, in a uniform steady magnetic field, the motion of particles parallel to the magnetic field and that perpendicular to the magnetic field is relatively independent, consisting of two parts: one is a uniform linear motion along the magnetic lines of force (the same as the case of no magnetic field); the other part is a whirling motion around the magnetic lines of force, which together make the charged particles spiral in a uniform steady magnetic field.[6] The direction of rotation of particles is opposite in the plane perpendicular to the magnetic field. Therefore, the transport properties of the plasma will be changed due to the presence of magnetic field.

Recently, nonextensive statistics has been widely applied to study the properties of the power-law distributed complex plasmas, including waves, instabilities and dust charging etc.[7-11] The transport processes of charged particles in the power-law distributed complex plasmas have also been studied. For example, some transport coefficients in the $\kappa$-distributed plasma were studied firstly using a Lorentz plasma model.[12] Later, the diffusion in a weakly ionized plasma with the $\kappa$-distributions was studied.[13] Most recently, some transport properties in the fully ionized plasma with $\kappa$-distributions were also studied.[14-16] However, the effects of magnetic field on the thermal conductivity in the $\kappa$-distributed plasmas have not been taken into consideration. In order to study the role of magnetic field in the astrophysical and space plasma, in this paper, we will consider the thermal conductivity and Dufour effects in the weakly ionized and magnetized plasma with the power-law velocity $\kappa$-distributions.

The paper is organized as follows. In Section 2, we introduce the generalized Boltzmann equation of transport for the weakly ionized, magnetized and power-law $\kappa$-distributed plasma and the first-order approximation of Chapman-Enskog expansion about the stationary-state velocity $\kappa$-distribution. In Section 3, we derive the effects of magnetic field on the transport properties including the thermal conductivity and Dufour effects. We also give graphic descriptions of the transport coefficients. Finally, in Section 4, we give the conclusion.

## 2 | THE GENERALIZED BOLTZMANN EQUATION OF TRANSPORT AND THE $\kappa$-DISTRIBUTIONS

In the weakly ionized plasma, because numbers of charged particles are much less than those of neutral molecules and/or atoms, only the elastic collisions between charged particles and neutral molecules or atoms are considered; while the collisions between charged particles are neglected. At the same time, the large angle deflections of charged particles are a result of collisions with neutral atoms or molecules rather than of multiple scattering with other charged particles. Based on the above facts of weakly ionized plasma, Boltzmann equation of transport together with further simplification of the collision integral are used to study transport properties of weakly ionized plasmas. It can be written as the generalized Boltzmann equation [13, 17-20],

$$\frac{\partial f_\alpha}{\partial t} + \mathbf{v} \cdot \nabla f_\alpha + \frac{Q_\alpha}{m_\alpha}\left[\mathbf{E} + c^{-1}\left(\mathbf{v}\times\mathbf{B}\right)\right]\cdot \nabla_v f_\alpha = C_\kappa(f_\alpha), \qquad (4)$$

where $f_\alpha \equiv f_\alpha(\mathbf{r}, \mathbf{v}, t,)$ is the single-particle velocity distribution function at time $t$, velocity $\mathbf{v}$ and position $\mathbf{r}$, and $\mathbf{E}$ is the electric field. The subscript $\alpha = e, i$ denotes electrons and ions, respectively. $Q_\alpha$ is charge of $\alpha$th component in the plasma. The term on the right-hand side, $C_\kappa(f_\alpha)$ is the $\kappa$-collisions term. In this sense, the stationary state is the $\kappa$-distribution when the system approaches to the $\kappa$-collision equilibrium.

Following the line of Refs.[13,18-20], the $\kappa$-collision term is considered as the generalized Krook model based on the $\kappa$-distribution. In the first-order approximation of Chapman-Enskog expansion for the velocity distribution functions in the weakly ionized, magnetized and $\kappa$-distributed plasma, we let the velocity distribution function to be $f_\alpha = f_{\kappa,\alpha}^{(0)} + f_{\kappa,\alpha}^{(1)}$, where the stationary-state distribution function is taken as the $\kappa$-distribution for $\alpha$th component, namely,

$$f_{\kappa,\alpha}^{(0)}(\mathbf{r},\mathbf{v}) = f_{\kappa,\alpha}(\mathbf{r},\mathbf{v}) = n_\alpha B_{\kappa,\alpha}\left(1 + A_{\kappa,\alpha}(\mathbf{v}-\mathbf{u})^2\right)^{-(\kappa_\alpha+1)} \qquad (5)$$

with

$$B_{\kappa,\alpha} = \left[\frac{2\pi k_B T_\alpha}{m_\alpha}\left(\kappa_\alpha - \frac{3}{2}\right)\right]^{-3/2} \frac{\Gamma(\kappa_\alpha+1)}{\Gamma(\kappa_\alpha-\frac{1}{2})}, \text{ and } A_{\kappa,\alpha} = \frac{m_\alpha}{(2\kappa_\alpha-3)k_B T_\alpha}.$$

And therefore, the transport equation (4) can be written as

$$\left\{\frac{\partial}{\partial t} + \mathbf{v}\cdot\frac{\partial}{\partial \mathbf{r}} + \frac{Q_\alpha}{m_\alpha}\left[\mathbf{E} + c^{-1}\mathbf{v}\times\mathbf{B}\right]\cdot\frac{\partial}{\partial \mathbf{v}}\right\}\left(f_{\kappa,\alpha} + f_{\kappa,\alpha}^{(1)}\right) = -\nu_\alpha f_{\kappa,\alpha}^{(1)}, \qquad (6)$$

where $\nu_\alpha$ is the mean collision frequency, $f_{\kappa,\alpha}^{(1)}$ is the first-order small disturbance about the stationary $\kappa$-distribution $f_{\kappa,\alpha}$. In general, the temperature and density in the nonequilibrium complex plasma are both considered to be spatially inhomogeneous and they can vary with time.

Usually, since transport processes are studied in a steady state so that $\partial f_\alpha/\partial t = 0$ and the first-order small disturbance satisfies $f_{\kappa,\alpha}^{(1)} \ll f_{q,\alpha}$, we can neglect $f_{\kappa,\alpha}^{(1)}$ on the left side of Eq. (6), and at the same time we find $(\mathbf{v}\times\mathbf{B})\cdot\nabla_v f_{\kappa,\alpha} = 0$. Thus Eq. (6) becomes

$$\mathbf{v}\cdot\nabla f_{\kappa,\alpha} + \frac{Q_\alpha \mathbf{E}}{m_\alpha}\cdot\nabla_v f_{\kappa,\alpha} + \frac{Q_\alpha}{m_\alpha c}(\mathbf{v}\times\mathbf{B})\cdot\nabla_v f_{\kappa,\alpha}^{(1)} = -\nu_\alpha f_{\kappa,\alpha}^{(1)}. \qquad (7)$$

For further simplification, we assume that $f_{\kappa,\alpha}^{(1)}$ is axisymmetric about the direction of current and is expressed [21] as

$$f_{\kappa,\alpha}^{(1)} = \frac{\mathbf{v}}{v}\cdot\mathbf{f}_{\kappa,\alpha}^{(1)}, \qquad (8)$$

where $\mathbf{f}_{\kappa,\alpha}^{(1)}$ is a vector along the direction of current and its velocity distribution is spherically symmetric. Thus we have that

$$\nabla_v f_{\kappa,\alpha}^{(1)} = \frac{\mathbf{v}\mathbf{v}}{v^2}\cdot\frac{\partial \mathbf{f}_{\kappa,\alpha}^{(1)}}{\partial v} + \left(\frac{\mathbf{I}}{v} - \frac{\mathbf{v}\mathbf{v}}{v^3}\right)\cdot\mathbf{f}_{\kappa,\alpha}^{(1)}, \qquad (9)$$

where $\mathbf{I}$ is a unit tensor. Using Eq. (9), Eq. (7) is changed as

$$\mathbf{v}\cdot\nabla f_{\kappa,\alpha}+\frac{Q_\alpha \mathbf{E}}{m_\alpha}\cdot\nabla_\mathbf{v} f_{\kappa,\alpha}+\frac{Q_\alpha}{m_\alpha c}\frac{\mathbf{v}}{v}\cdot\left(\mathbf{B}\times\mathbf{f}_{\kappa,\alpha}^{(1)}\right)=-\nu_\alpha \frac{\mathbf{v}}{v}\cdot\mathbf{f}_{\kappa,\alpha}^{(1)}\ ,\qquad(10)$$

where we have used that $\mathbf{v}\times\mathbf{B}\cdot\mathbf{vv}=0$ and $\mathbf{v}\times\mathbf{B}\cdot\mathbf{I}\cdot\mathbf{f}_{\kappa,\alpha}^{(1)}=\mathbf{v}\cdot\left(\mathbf{B}\times\mathbf{f}_{\kappa,\alpha}^{(1)}\right)$.

Because the transport of the particles parallel to the magnetic field is the same as that without magnetic field, in order to study the effect of magnetic field, we here only consider the parts of the transport quantities perpendicular to the magnetic field. We multiply Eq. (10) by $3\mathbf{v}/v$ and then integrate it with respect to the solid angle $\Omega$. And considering that $\frac{1}{4\pi}\iint\frac{\mathbf{vv}}{v^2}d\Omega=\frac{1}{3}\mathbf{I}$, we have that

$$v\nabla_\perp f_{\kappa,\alpha}+\frac{Q_\alpha \mathbf{E}_\perp}{m_\alpha}\cdot\frac{\partial f_{\kappa,\alpha}}{\partial v}-\omega_{B,\alpha}\left(\hat{\mathbf{B}}\times\mathbf{f}_{\kappa,\alpha}^{(1)}\right)=-\nu_\alpha \mathbf{f}_{\kappa,\alpha}^{(1)}\ ,\qquad(11)$$

where $\omega_{B,\alpha}=|Q_\alpha|B/m_\alpha c$ is the cyclotron frequency of charged particles, $\hat{\mathbf{B}}=\mathbf{B}/B$ is the unit vector along the direction of magnetic field.

A vector product of $\hat{\mathbf{B}}$ is made on both sides of Eq. (11), we obtain

$$v\left(\hat{\mathbf{B}}\times\nabla_\perp f_{\kappa,\alpha}\right)+\frac{Q_\alpha}{m_\alpha}\frac{\partial f_{\kappa,\alpha}}{\partial v}\left(\hat{\mathbf{B}}\times\mathbf{E}_\perp\right)+\omega_{B,\alpha}\mathbf{f}_{\kappa,\alpha}^{(1)}=-\nu_\alpha\left(\hat{\mathbf{B}}\times\mathbf{f}_{\kappa,\alpha}^{(1)}\right).\qquad(12)$$

We solve $\mathbf{f}_{\kappa,\alpha}^{(1)}$ for Eq. (11) and Eq. (12), and find that

$$\mathbf{f}_{\kappa,\alpha}^{(1)}=\frac{1}{\nu_\alpha^2+\omega_B^2}\left[-\nu_\alpha v\nabla_\perp f_{\kappa,\alpha}-\omega_{B,\alpha}v\left(\hat{\mathbf{B}}\times\nabla_\perp f_{\kappa,\alpha}\right)-\left(\omega_{B,\alpha}\hat{\mathbf{B}}\times\mathbf{E}_\perp+\nu_\alpha\mathbf{E}_\perp\right)\frac{Q_\alpha}{m_\alpha}\frac{\partial f_{\kappa,\alpha}}{\partial v}\right].\qquad(13)$$

Therefore, in the first-order approximation of Chapman-Enskog expansion, from Eq. (8) we obtain the velocity distribution functions of charged particles in the weakly ionized, magnetized and power-law $\kappa$-distributed plasma,

$$f_\alpha=f_{\kappa,\alpha}-\frac{\mathbf{v}}{\nu_\alpha^2+\omega_B^2}\cdot\left[\omega_{B,\alpha}\left(\hat{\mathbf{B}}\times\nabla_\perp f_{\kappa,\alpha}\right)+\nu_\alpha\nabla_\perp f_{\kappa,\alpha}+\left(\omega_{B,\alpha}\hat{\mathbf{B}}\times\mathbf{E}_\perp+\nu_\alpha\mathbf{E}_\perp\right)\frac{Q_\alpha}{vm_\alpha}\frac{\partial f_{\kappa,\alpha}}{\partial v}\right].\qquad(14)$$

## 3 | THERMAL CONDUCTIVITY AND DUFOUR EFFECTS

In a nonequilibrium plasma system, the heat flux can be generated not only by the temperature gradient but also by the density gradient, where the part driven by the temperature gradient is the heat conduction phenomenon and the part driven by the density gradient is the Dufour effects. The macroscopic law for the thermodynamic "flux" and the thermal dynamic "force" is expressed [22] as

$$\mathbf{q}=-\left(K+\frac{D_\rho}{2T}\right)\nabla T-\frac{D_\rho}{2\rho}\nabla\rho\ ,\qquad(15)$$

where $\mathbf{q}$ is the heat flux vector of the plasma, $\rho$ is the mass density, $K$ is the heat conductivity and $D_\rho$ is the Dufour coefficient. If the mass density is expressed using the particle number density $n$ as $\rho=mn$, then the heat flux vector becomes

$$\mathbf{q}=-\left(K+\frac{D_\rho}{2T}\right)\nabla T-\frac{D_\rho}{2n}\nabla n\ ,\qquad(16)$$

In the magnetized nonequilibrium plasma, where the transport process is driven by a force $\mathbf{s}$, the transport coefficient tensor $\psi$ can be expressed [23,24] as

$$\psi\cdot\mathbf{s}=\psi_\parallel\hat{\mathbf{B}}\left(\hat{\mathbf{B}}\cdot\mathbf{s}\right)+\psi_\perp\hat{\mathbf{B}}\times\left(\mathbf{s}\times\hat{\mathbf{B}}\right)+\psi_\wedge\hat{\mathbf{B}}\times\mathbf{s}\ .\qquad(17)$$

The unit vector $\hat{\mathbf{B}}$ provides a reference for describing transport phenomena. The transport process can be divided into three directions: [23] $\psi_\parallel \hat{\mathbf{B}}(\hat{\mathbf{B}}\cdot\mathbf{s})$ is the transport in the same direction as the magnetic field, proportional to **s** component parallel to $\hat{\mathbf{B}}$; $\psi_\perp \hat{\mathbf{B}}\times(\mathbf{s}\times\hat{\mathbf{B}})$ is the transport perpendicular to the magnetic field, proportional to **s** component perpendicular to $\hat{\mathbf{B}}$; $\psi_\wedge \hat{\mathbf{B}}\times\mathbf{s}$ is the transport perpendicular to the plane composed of $\hat{\mathbf{B}}$ and **s**.

Here we do not consider the transport parallel to the direction of magnetic field because it is the same as that without magnetic field. In the magnetized nonequlibrium plasma, the macroscopic law of heat conduction process of $\alpha$th component in the magnetized plasma is expressed as

$$\mathbf{q}_\alpha = -\left(K_{\perp,\alpha}+\frac{D_{\rho\perp,\alpha}}{2T_\alpha}\right)\nabla_\perp T_\alpha - \left(K_{\wedge,\alpha}+\frac{D_{\rho\wedge,\alpha}}{2T_\alpha}\right)\hat{\mathbf{B}}\times\nabla T_\alpha - \frac{D_{\rho\perp,\alpha}}{2n_\alpha}\nabla_\perp n_\alpha - \frac{D_{\rho\wedge,\alpha}}{2n_\alpha}\hat{\mathbf{B}}\times\nabla n_\alpha. \qquad (18)$$

where we denote that

$$\nabla_\perp T_\alpha \equiv \hat{\mathbf{B}}\times\left(\nabla T_\alpha\times\hat{\mathbf{B}}\right) \text{ and } \nabla_\perp n_\alpha \equiv \hat{\mathbf{B}}\times\left(\nabla n_\alpha\times\hat{\mathbf{B}}\right),$$

and the transport coefficients $K_{\perp,\alpha}$, $D_{\rho\perp,\alpha}$, $K_{\wedge,\alpha}$ and $D_{\rho\wedge,\alpha}$ are respectively the transverse conductivity, the transverse Dufour coefficient, the Righi-Leduc coefficient and the Hall Dufour coefficient.

The heat flux vector for the $\alpha$th component of the plasma is defined [25] as

$$\mathbf{q}_\alpha = \frac{1}{2}m_\alpha \int (\mathbf{v}-\mathbf{u})^2 (\mathbf{v}-\mathbf{u}) f_\alpha d\mathbf{v}, \qquad (19)$$

Therefore, if the plasma velocity distributions are considered as the $\kappa$-distributions, by substituting Eq. (14) into Eq. (19) and omitting the terms containing the electric field, we obtain that

$$\mathbf{q}_\alpha = \frac{1}{2}m_\alpha \int d\mathbf{v}(\mathbf{v}-\mathbf{u})^3 \left\{ f_{\kappa,\alpha} - \left[\frac{\mathbf{v}_\alpha}{v_\alpha^2+\omega_{B,\alpha}^2}\cdot\left(\omega_{B,\alpha}\hat{\mathbf{B}}\times\nabla_\perp f_{\kappa,\alpha}+v_\alpha\nabla_\perp f_{\kappa,\alpha}\right)\right]\right\}, \qquad (20)$$

where the following integral is zero,

$$\int d\mathbf{v}(\mathbf{v}-\mathbf{u})^3 f_{\kappa,\alpha} = n_\alpha B_{\kappa,\alpha}\int \mathbf{w}^2 \mathbf{w}\left[1+A_{\kappa,\alpha}\mathbf{w}^2\right]^{-(\kappa_\alpha+1)} d\mathbf{w} = 0,$$

with **w** = **v**-**u**, because the integrand function is an odd function of the velocity. Thus Eq. (20) becomes

$$\mathbf{q}_\alpha = \frac{1}{2}m_\alpha \int d\mathbf{v}(\mathbf{v}-\mathbf{u})^3 \left[\frac{-\mathbf{v}_\alpha}{v_\alpha^2+\omega_{B,\alpha}^2}\cdot\left(\omega_{B,\alpha}\hat{\mathbf{B}}\times\nabla_\perp f_{\kappa,\alpha}+v_\alpha\nabla_\perp f_{\kappa,\alpha}\right)\right]. \qquad (21)$$

After the integrals in Eq. (21) are completed (see Appendix), we can derive that

$$\mathbf{q}_\alpha = -\frac{5(2\kappa_\alpha-3)k_B^2 T_\alpha^2 v_\alpha}{2m_\alpha(2\kappa_\alpha-5)(v_\alpha^2+\omega_{B,\alpha}^2)}\nabla_\perp n_\alpha - \frac{5(2\kappa_\alpha-3)v_\alpha n_\alpha k_B^2 T_\alpha}{m_\alpha(2\kappa_\alpha-5)(v_\alpha^2+\omega_{B,\alpha}^2)}\nabla_\perp T_\alpha$$

$$-\frac{5(2\kappa_\alpha-3)\omega_{B,\alpha}k_B^2 T_\alpha^2}{2m_\alpha(2\kappa_\alpha-5)(v_\alpha^2+\omega_{B,\alpha}^2)}(\hat{\mathbf{B}}\times\nabla_\perp n_\alpha) - \frac{5(2\kappa_\alpha-3)n_\alpha k_B^2 T_\alpha \omega_{B,\alpha}}{m_\alpha(2\kappa_\alpha-5)(v_\alpha^2+\omega_{B,\alpha}^2)}(\hat{\mathbf{B}}\times\nabla_\perp T_\alpha). \qquad (22)$$

Comparing with the thermodynamic equation (18), we can find the transverse thermal conductivity, the Righi-Leduc coefficient, the transverse Dufour coefficient and the Hall Dufour coefficient in the plasma with power-law $\kappa$–distributions, respectively,

$$K_{\perp,\alpha}^\kappa = \frac{5v_\alpha n_\alpha k_B^2 T_\alpha}{2m_\alpha(v_\alpha^2+\omega_{B,\alpha}^2)}\frac{(2\kappa_\alpha-3)}{(2\kappa_\alpha-5)}, \qquad (23)$$

$$K^{\kappa}_{\wedge,\alpha} = \frac{5n_\alpha k_B^2 T_\alpha \omega_{B,\alpha}}{2m_\alpha \left(v_\alpha^2 + \omega_{B,\alpha}^2\right)} \frac{(2\kappa_\alpha - 3)}{(2\kappa_\alpha - 5)}, \quad (24)$$

$$D^{\kappa}_{\rho\perp,\alpha} = \frac{5v_\alpha n_\alpha k_B^2 T_\alpha^2}{m_\alpha \left(v_\alpha^2 + \omega_{B,\alpha}^2\right)} \frac{(2\kappa_\alpha - 3)}{(2\kappa_\alpha - 5)}, \quad (25)$$

$$D^{\kappa}_{\rho\wedge,\alpha} = \frac{5n_\alpha k_B^2 T_\alpha^2 \omega_{B,\alpha}}{m_\alpha \left(v_\alpha^2 + \omega_{B,\alpha}^2\right)} \frac{(2\kappa_\alpha - 3)}{(2\kappa_\alpha - 5)}, \quad (26)$$

Eqs. (23)-(26) show that these new transport coefficients depend strongly on the $\kappa$-parameter and the magnetic field. When the magnetic field is equal to zero, the Righi-Leduc coefficient (24) and the Hall Dufour coefficient (26) disappear, and the transverse thermal conductivity (23) and the Dufour coefficient (25) get the shape for the plasma without magnetic field.[20] When we take the parameter $\kappa \to \infty$, they recover to the standard forms in the plasma with a Maxwellian velocity distribution,[26-28] namely,

$$K^{\infty}_{\perp,\alpha} = \frac{5n_\alpha v_\alpha k_B^2 T_\alpha}{2m_\alpha \left(v_\alpha^2 + \omega_{B,\alpha}^2\right)}, \quad K^{\infty}_{\wedge,\alpha} = -\frac{5n_\alpha k_B^2 T_\alpha \omega_{B,\alpha}}{2m_\alpha \left(v_\alpha^2 + \omega_{B,\alpha}^2\right)}, \quad (27)$$

$$D^{\infty}_{\rho\perp,\alpha} = \frac{5v_\alpha n_\alpha k_B^2 T_\alpha^2}{m_\alpha \left(v_\alpha^2 + \omega_{B,\alpha}^2\right)}, \quad D^{\infty}_{\rho\wedge,\alpha} = -\frac{5n_\alpha k_B^2 T_\alpha^2 \omega_{B,\alpha}}{m_\alpha \left(v_\alpha^2 + \omega_{B,\alpha}^2\right)}. \quad (28)$$

In order to show more clearly the roles of the $\kappa$-parameter as well as of the magnetic field in the thermal conductivity and Dufour effects, we relate the coefficients of Eqs. (23)-(26) to the thermal conductivity and Dufour coefficient in the unmagnetized Maxwellian plasma,[20, 26, 29]

$$K^{\infty}_\alpha = \frac{5n_\alpha k_B^2 T_\alpha}{2m_\alpha v_\alpha}, \quad D^{\infty}_{\rho,\alpha} = \frac{5n_\alpha k_B^2 T_\alpha^2}{m_\alpha v_\alpha}. \quad (29)$$

And then, from Eqs.(23)-(26) and (29) we get that

$$\frac{K^{\kappa}_{\perp,\alpha}}{K^{\infty}_\alpha} = \frac{D^{\kappa}_{\rho\perp,\alpha}}{D^{\infty}_{\rho,\alpha}} = \frac{v_\alpha^2}{\left(v_\alpha^2 + \omega_B^2\right)} \frac{(2\kappa_\alpha - 3)}{(2\kappa_\alpha - 5)}, \quad (30)$$

$$\frac{K^{\kappa}_{\wedge,\alpha}}{K^{\infty}_\alpha} = \frac{D^{\kappa}_{\rho\wedge,\alpha}}{D^{\infty}_{\rho,\alpha}} = \frac{\omega_B v_\alpha}{\left(v_\alpha^2 + \omega_B^2\right)} \frac{(2\kappa_\alpha - 3)}{(2\kappa_\alpha - 5)}. \quad (31)$$

On the basis of the above equations, Eqs.(29)-(31), in Figs.1-2 we show the roles of magnetic field in the transport coefficients, which are relative to those in the case of a Maxwellian distribution and no magnetic field, and the calculations are made for three different values of the $\kappa$-parameters, $\kappa_\alpha = 4$, 5 and $\infty$, where $K^{\kappa}_{\perp,\alpha}/K^{\infty}_\alpha = D^{\kappa}_{\rho\perp,\alpha}/D^{\infty}_{\rho,\alpha}$ and $K^{\kappa}_{\wedge,\alpha}/K^{\infty}_\alpha = D^{\kappa}_{\rho\wedge,\alpha}/D^{\infty}_{\rho,\alpha}$ are respectively as the ordinate axis and $\omega_{Be}/v_\alpha$ is as the abscissa axis in the figures.

In Figure 1, we show for the three different $\kappa$-parameters, that the transverse thermal conductivity and the transverse Dufour coefficient both decrease monotonously with the increases of magnetic field, and they also both decrease with the increase of the $\kappa$-parameters. And we find that the thermal conductivity and the Dufour coefficient in the $\kappa$-distributed plasma are both generally greater than those in the Maxwell-distributed plasma.

In Figure 2, we show that, for the three different $\kappa$-parameters in the plasma, with the increases of the magnetic field, the Righi-Leduc coefficient and the Hall Dufour coefficient both increase at first, then reach the maximum, and finally decrease. And we find that in the $\kappa$-distributed plasma, the two coefficients are both generally greater than those in the Maxwell-distributed plasma.

In all the four transport coefficients, the $\kappa$-parameter $\kappa \neq \infty$ plays significant roles.

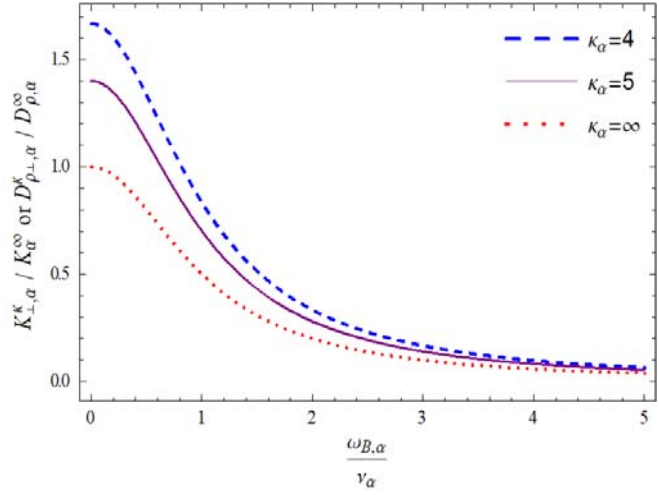

**FIGURE 1** The thermal conductivity and the Dufour coefficient as a function of the magnetic field effect for three different $\kappa$-parameters.

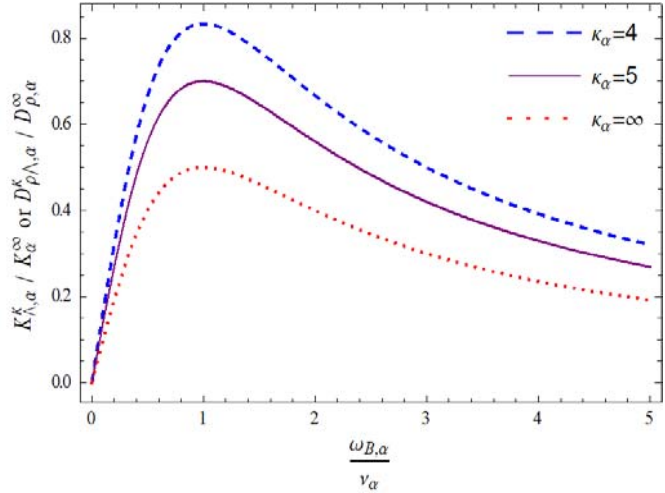

**FIGURE 2** The Righi-Leduc coefficient and the Hall Dufour coefficient as a function of the magnetic field effect for three different $\kappa$-parameters.

## 4 | CONCLUSIONS

In conclusion, by using the generalized Boltzmann equation of transport and in the first-order approximation of Chapman-Enskog expansion about the stationary-state $\kappa$-distribution of the plasma, we have studied the effect of steady magnetic field on the thermal conductivity and Dufour effects of weakly ionized and magnetized plasma with the velocity $\kappa$-distribution. Using the macroscopic thermodynamic laws with the magnetic field and the corresponding microscopic expressions, we have derived the transverse thermal conductivity, the Righi-Leduc coefficient, the transverse Dufour coefficient and the Hall Dufour coefficient in the plasma, which are given by Eqs. (23)-(26), respectively.

It is shown that the magnetic field also has a significant effect on all these transport coefficients in the $\kappa$-distributed plasma. And the transverse thermal conductivity, the Righi-Leduc coefficient, the transverse Dufour coefficient and the Hall Dufour coefficient depend strongly on the $\kappa$-parameter $\kappa \neq \infty$ of the plasma. In other words, in the $\kappa$-distributed plasma, the thermal conductivity tensor and Dufour coefficient tensor are significantly different from those in the Maxwell-distributed plasma.

For three different $\kappa$-parameters in the plasma, we have given graphic analyses of the thermal conductivity, the Righi-Leduc coefficient and the Dufour coefficients, which are shown clearly as a function of the magnetic field. We find the property that the thermal conductivity, the Dufour coefficient, the Righi-Leduc coefficient and the Hall Dufour coefficient in the $\kappa$-distributed plasma are generally greater than those in the Maxwell-distributed plasma.

In this paper, as usual, the mean collision frequency $\nu_\alpha$ has been assumed to be a constant. However, if the mean collision frequency in the velocity $\kappa$-distributed plasma depends significantly on the $\kappa$-parameter, then it will modify the properties of the transport coefficients in the plasma affected by the magnetic field.

**ACKNOWLEDGMENTS**

This work is supported by the National Natural Science Foundation of China under Grant No. 11775156.

**APPENDIX**

In Eq. (21), the two integrals are, respectively, that

$$\int (\mathbf{v}-\mathbf{u})^3 \left[\mathbf{v}\cdot\nabla_\perp f_{\kappa,\alpha}\right] d\mathbf{v} = \int (\mathbf{v}-\mathbf{u})^3 \mathbf{v} \cdot \left\{\frac{\nabla_\perp n_\alpha}{n_\alpha} + \left[(\kappa_\alpha+1)\left(1+\frac{1}{(\mathbf{v}-\mathbf{u})^2 A_{\kappa,\alpha}}\right)^{-1} - \frac{3}{2}\right]\frac{\nabla_\perp T_\alpha}{T_\alpha}\right\} f_{\kappa,\alpha} d\mathbf{v}_\alpha$$

$$= \frac{\nabla_\perp T_\alpha}{T_\alpha}\int (\mathbf{v}-\mathbf{u})^3 \mathbf{v} \left[(\kappa_\alpha+1)\left(1+\frac{1}{(\mathbf{v}-\mathbf{u})^2 A_{\kappa,\alpha}}\right)^{-1} - \frac{3}{2}\right] f_{\kappa,\alpha} d\mathbf{v} + \frac{\nabla_\perp n_\alpha}{n_\alpha}\int (\mathbf{v}-\mathbf{u})^3 \mathbf{v} f_{\kappa,\alpha} d\mathbf{v}, \quad (A.1)$$

where the integrals are calculated as

$$\int \mathbf{v}(\mathbf{v}-\mathbf{u})^3 f_{\kappa,\alpha} d\mathbf{v} = n_\alpha B_{\kappa,\alpha} \int \mathbf{v}(\mathbf{v}-\mathbf{u})^3 \left[1+A_{\kappa,\alpha}(\mathbf{v}-\mathbf{u})^2\right]^{-(\kappa+1)} d\mathbf{v}$$

$$= n_\alpha B_{\kappa,\alpha} \int \mathbf{w}^3(\mathbf{w}+\mathbf{u})\left[1+A_{\kappa,\alpha}\mathbf{w}^2\right]^{-(\kappa+1)} d\mathbf{w}$$

$$= n_\alpha B_{\kappa,\alpha} \frac{4\pi}{3}\int_0^\infty w^6\left[1+A_{\kappa,\alpha}\mathbf{w}^2\right]^{-(\kappa+1)} dw$$

$$= n_\alpha B_{\kappa,\alpha} \frac{5\pi\sqrt{\pi}}{4 A_{\kappa,\alpha}^{7/2}} \frac{\Gamma(\kappa_\alpha-\tfrac{5}{2})}{\Gamma(1+\kappa_\alpha)} = \frac{5(2\kappa_\alpha-3)n_\alpha k_B^2 T_\alpha^2}{(2\kappa_\alpha-5)m_\alpha^2}, \quad (A.2)$$

$$\int (\mathbf{v}-\mathbf{u})^3 \mathbf{v}\left[(\kappa_\alpha+1)\left(1+\frac{1}{(\mathbf{v}-\mathbf{u})^2 A_{\kappa,\alpha}}\right)^{-1} - \frac{3}{2}\right] f_{\kappa,\alpha} d\mathbf{v}$$

$$= (\kappa_\alpha+1)A_{\kappa,\alpha} n_\alpha B_{\kappa,\alpha} \int (\mathbf{v}-\mathbf{u})^5 \mathbf{v}\left[1+A_{\kappa,\alpha}(\mathbf{v}-\mathbf{u})^2\right]^{-(\kappa+2)} d\mathbf{v} - \frac{3}{2}\int (\mathbf{v}-\mathbf{u})^3 \mathbf{v} f_{\kappa,\alpha} d\mathbf{v}$$

$$= (\kappa_\alpha + 1) A_{\kappa,\alpha} n_\alpha B_{\kappa,\alpha} \int \mathbf{w}^5 (\mathbf{u}+\mathbf{w}) \left[1 + A_{\kappa,\alpha} \mathbf{w}^2\right]^{-(\kappa+2)} d\mathbf{w} - \frac{3}{2} \int \mathbf{w}^3 (\mathbf{u}+\mathbf{w}) f_{\kappa,\alpha} d\mathbf{w}$$

$$= \frac{4\pi}{3} (\kappa_\alpha + 1) A_{\kappa,\alpha} n_\alpha B_{\kappa,\alpha} \int_0^\infty w^8 \left[1 + A_{\kappa,\alpha} w^2\right]^{-(\kappa+2)} dw - \frac{15(2\kappa_\alpha - 3) n_\alpha k_B^2 T_\alpha^2}{2(2\kappa_\alpha - 5) m_\alpha^2}$$

$$= \frac{35\pi\sqrt{\pi}}{8 A_{\kappa,\alpha}^{7/2}} (\kappa_\alpha + 1) n_\alpha B_{\kappa,\alpha} \frac{\Gamma(\kappa_\alpha - \frac{5}{2})}{\Gamma(\kappa+2)} - \frac{15(2\kappa_\alpha - 3) n_\alpha k_B^2 T_\alpha^2}{2(2\kappa_\alpha - 5) m_\alpha^2}$$

$$= \frac{10(2\kappa_\alpha - 3) n_\alpha k_B^2 T_\alpha^2}{(2\kappa_\alpha - 5) m_\alpha^2}. \tag{A.3}$$

Thus Eq. (A1) becomes

$$\int (\mathbf{v}_\alpha - \mathbf{u}_\alpha)^2 (\mathbf{v}_\alpha - \mathbf{u}_\alpha) \left[\mathbf{v}_\alpha \cdot \nabla_\perp f_{\kappa,\alpha}\right] d\mathbf{v}_\alpha$$
$$= \frac{5(2\kappa_\alpha - 3) k_B^2 T_\alpha^2}{(2\kappa_\alpha - 5) m_\alpha^2} \nabla_\perp n_\alpha + \frac{10(2\kappa_\alpha - 3) n_\alpha k_B^2 T_\alpha}{(2\kappa_\alpha - 5) m_\alpha^2} \nabla_\perp T_\alpha. \tag{A.4}$$

In the same way, we can calculate that

$$\int (\mathbf{v}_\alpha - \mathbf{u}_\alpha)^2 (\mathbf{v}_\alpha - \mathbf{u}_\alpha) \left[\mathbf{v}_\alpha \cdot (\hat{\mathbf{B}} \times \nabla_\perp f_{\kappa,\alpha})\right] d\mathbf{v}_\alpha$$
$$= \frac{5(2\kappa_\alpha - 3) k_B^2 T_\alpha^2}{(2\kappa_\alpha - 5) m_\alpha^2} (\hat{\mathbf{B}} \times \nabla_\perp n_\alpha) + \frac{10(2\kappa_\alpha - 3) n_\alpha k_B^2 T_\alpha}{(2\kappa_\alpha - 5) m_\alpha^2} (\hat{\mathbf{B}} \times \nabla_\perp T_\alpha) \tag{A.5}$$

Substituting Eqs. (A4) and (A5) into Eq. (21), we can derive Eq. (22).